\documentclass[12pt]{iopart}
\usepackage{harvard}
\usepackage{graphicx}
\usepackage{subfig}
\newcommand{\be}{\begin{equation}}
\newcommand{\ee}{\end{equation}}
\newcommand{\Mc}{{\cal M}}
\newcommand{\Ms}{M_{\odot}}

\usepackage[usenames,dvips]{color}

\begin{document}

\title[Parameter estimation of gravitational-wave signals with spin effects]{The effects of LIGO detector noise on a 15-dimensional Markov-chain Monte-Carlo analysis of gravitational-wave signals}

\author{V. Raymond$^{1}$, M.V.\ van der Sluys$^{2}$, I.\ Mandel$^{1}$, 
V.\ Kalogera$^{1}$, C.\ R\"{o}ver$^{3}$, N.\ Christensen$^{4}$}
% \author{Vivien Raymond$^{1}$ for the LSC}

\address{$^1$ Dept.\ of Physics \& Astronomy, Northwestern University, 2131 Tech Drive,  Evanston IL, 60208,  USA} 
\address{$^2$ Dept.\ of Physics, University of Alberta, 11322 -- 89 Avenue, Edmonton AB, T6G 2G7, Canada}
\address{$^3$ Max-Planck-Institut f\"{u}r Gravitationsphysik, Callinstra\ss{}e 38, 30167 Hannover, Germany}
\address{$^4$ Physics \& Astronomy Dept., Carleton College, One North College Street, Northfield MN, 55057, USA}

\ead{vivien@u.northwestern.edu}

\begin{abstract}
  Gravitational-wave signals from inspirals of binary compact objects (black holes and neutron stars) are primary targets of the ongoing searches 
  by ground-based gravitational-wave (GW) interferometers (LIGO, Virgo, and GEO-600). 
  We present parameter-estimation results from our Markov-chain Monte-Carlo code \textsc{SPINspiral} on signals from binaries with precessing spins. 
  Two data sets are created by injecting simulated GW signals into either synthetic Gaussian noise or into LIGO detector data. We
  compute the 15-dimensional probability-density functions (PDFs) for both data sets, as well as for a data set containing LIGO data with
  a known, loud artefact (``glitch''). 
  We show that the analysis of the signal in detector noise yields accuracies similar to those obtained using simulated Gaussian noise.  We also find 
  that while the Markov chains from the glitch do not converge, the PDFs would look consistent 
  with a GW signal present in the data. While our parameter-estimation results are encouraging, further investigations into how to differentiate an actual GW signal from noise are necessary.
  %These results warrant further systematic investigations.
\end{abstract}

%\noindent \emph{Keywords}: Binaries: close, Gamma rays: bursts, Gravitational waves, Relativity, Markov-chain Monte Carlo
\maketitle

\section{Introduction}
\label{sec:intro}

Among the sources of gravitational waves (GWs), inspiralling binary systems of compact objects, neutron stars (NSs) and/or black holes (BHs) 
in the mass range $\sim 1\,\Ms - 100\,\Ms$ stand out as likely to be detected and relatively easy to model. 
For ground-based laser interferometers currently in operation 
\cite{2002gr.qc.....4090C}, LIGO \cite{2009NJPh...11g3032A}, Virgo \cite{2008CQGra..25r4001A} and GEO-600 \cite{2004CQGra..21S.417W},
the current detection-rate estimates for BH-NS binaries range from $2\times10^{-4}$ to $0.2$\,yr$^{-1}$ for first-generation 
instruments \citeaffixed{2008ApJ...672..479O,2010arXiv1003.2480L}{\emph{e.g.}}. 
Although the estimates are quite uncertain, detection rates are expected to increase %by factors of about $\sim 10$ 
with the upgrade 
%to Enhanced LIGO/Virgo (now--2011) and by $\sim 10^3$ with Advanced LIGO/Virgo (2015--2017). 
to Enhanced LIGO/Virgo, up to $\sim 40$\,yr$^{-1}$ with Advanced LIGO/Virgo. 

The detection of a gravitational-wave event is challenging and will be a rewarding achievement by itself. After such a detection, measurement of source properties holds major 
promise for improving our astrophysical understanding and requires reliable methods for parameter estimation. 
This is a complicated problem, because of the large number of parameters ($15$ for spinning compact objects in a quasi-circular orbit) 
and the degeneracies between them \cite{2009CQGra..26k4007R}, the significant amount of structure in the parameter space, and the particularities of the detector noise. 

%In the case of low-mass-ratio binaries (\emph{e.g.}\ BH-NS), these issues are amplified 
%for significant spin magnitudes and large misalignments between the spins and the orbital angular momentum
%\cite{1994PhRvD..49.6274A,2003PhRvD..67d2003G,2003PhRvD..67j4025B}. However, the presence of  
%spins improves parameter estimation through the signal modulations, although still  
%presenting us with a considerable computational challenge. This was highlighted  
%in the context of LISA observations \citeaffixed{2004PhRvD..70d2001V,2006PhRvD..74l2001L}{see}, and in our first study devoted to ground-based observations \cite{2008ApJ...688L..61V}.

In this paper we use an example to illustrate the capabilities of our Markov-chain Monte-Carlo (MCMC) algorithm \textsc{SPINspiral} \cite{2008CQGra..25r4011V} 
for parameter estimation of binary inspirals with two spinning components, using ground-based GW interferometers. 
In these proceedings we focus on the effects of using LIGO detector data versus synthetic Gaussian noise.
Earlier studies \citeaffixed{1994PhRvD..49.1723J,1994PhRvD..49.2658C,1995PhRvD..52..848P,2007CQGra..24.1089V}{\emph{e.g.}}
computed the potential accuracy of parameter estimation (\emph{e.g.}\ by using the Fisher matrix), 
but without performing a parameter estimation in practice.
%%Such methods were usually applied to objects
%%with zero or (anti)aligned spin and only work when the parameter values are already known and the 
%%posterior distribution is unimodal.
%%(see the end of Sect.\,\ref{sec:results} for a discussion).
Also,
\citename{2006CQGra..23.4895R} \citeyear{2006CQGra..23.4895R,2007PhRvD..75f2004R},  \citename{2009arXiv0911.3820V} \citeyear{2008PhRvD..78b2001V,2008CQGra..25r4010V,2009arXiv0911.3820V}
explored parameter estimation for binaries without spins, described by nine parameters.
%We focus on BH-NS binaries, 
%which can exhibit significant coupling between the orbital angular momentum and the BH spin, mainly because of 

%the high mass ratio \cite{1994PhRvD..49.6274A}, while at the same time we are justified to ignore the NS spin, 
%leading to a 12-dimensional parameter space.
We present the gravitational-wave template used for this study in section~\,\ref{sec:GW}, and the Bayesian framework we employ here in section~\,\ref{sec:methods}.
In section~\ref{sec:data} we describe the three data sets that we analyse in this study; a simulated GW signal injected into synthetic Gaussian noise,
a GW signal injected into LIGO detector data and a raw LIGO data set containing a known artefact of terrestrial origin (``glitch'').
We describe the details of the MCMC simulations in section~\ref{sec:runs}.  The analyses of the first two data sets are compared in section~\ref{sec:subresults},
and we present our results on the glitch in section~\,\ref{sec:glitch}.

%%Following our previous study \cite{2008ApJ...688L..61V}, we analyze the degeneracy in the sky-position determination when data from only two detectors are used.
%In our previous study \cite{2008ApJ...688L..61V}, we showed the accuracy obtained in sky-position determination 
%using data from a two-detector network, where a degeneracy in the sky position exists, and from a three-detector network, where the degeneracy is broken.
%Following this work, we here analyze in further detail the degeneracy which is present when data from only two detectors are used.
%In section~\ref{sec:skyring} we show that the degeneracy in the sky 
%position is reduced but not lifted when a significant spin is present ($a_\mathrm{spin}\ge0.5$), and that a 
%sufficient angle between spin and orbital angular momentum can break such a degeneracy 
%($\theta_{SL}\ge55^{\circ}$). In this study, we demonstrate that these degeneracies are due to 
%a high degree of similarity between signals from sources with significantly different parameter 
%sets, while we made sure the observed effects are real and
%not in fact artifacts due to potential errors in our analysis
%methodology.

%In section~\,\ref{sec:comparison}, we also demonstrate that the inclusion of 
%higher post-Newtonian orders in the waveform can improve the accuracy of intrinsic-parameter estimation.  Meanwhile, we find 
%that the additional structure in the parameter space of higher-order waveforms lowers the sampling efficiency of the MCMC and requires improvements to the sampling scheme.

\section{Gravitational-wave signal and observables}
\label{sec:GW}

%The signal injected was the inspiral of a 1.4 \Ms$ - 1.12 solar mass compact object binary, simulated by a 3pN expansion. 

We analyse the signal produced during the inspiral phase of two compact objects of masses $M_{1,2}$ in quasi-circular orbit. 
We focus on a black-hole binary system with $M_1 = 10\,\Ms$ and $M_2 = 1.4\,\Ms$, where unlike in some of our previous studies \citeaffixed{2008ApJ...688L..61V}{\emph{e.g.}}, 
we do not ignore the second spin to explore the single spin approximation.
During the orbital inspiral, the general-relativistic spin-orbit and spin-spin coupling (dragging of inertial frames) cause the binary's orbital plane to precess 
and introduce amplitude and phase modulations of the observed gravitational-wave signal \cite{1994PhRvD..49.6274A}.

A circular binary inspiral with both compact objects spinning is described by a 15-dimensional parameter  
vector $\vec{\lambda} \in \Lambda$. Our choice of independent parameters with respect to a fixed geocentric coordinate system is:
%\vec{\lambda} = \{\Mc,\eta,d_\mathrm{L},t_\mathrm{c},\phi_\mathrm{c},\alpha,\delta,\iota,\psi,a_\mathrm{spin1},\theta_\mathrm{spin1},\phi_\mathrm{spin1},a_\mathrm{spin2},\theta_\mathrm{spin2},\phi_\mathrm{spin2}\},
%%\vec{\lambda}\!=\!\{\!\Mc\!,\eta\!,\mathrm{RA}\!,\cos\mathrm{Dec}\!,\cos\theta_{J_0}\!,\phi_{J_0}\!,\log{d_\mathrm{L}}\!,a_\mathrm{spin}\!,\cos\theta_{SL}\!, \phi_\mathrm{c}\!,\alpha_\mathrm{c}\!, t_\mathrm{c}\!\}\!,
\begin{eqnarray}
  \vec{\lambda} =& \{\Mc,\eta,\log{d_\mathrm{L}},t_\mathrm{c},\phi_\mathrm{c},\alpha,\cos\delta,\sin{\iota},\psi, \nonumber \\
  &a_\mathrm{spin1},\cos\theta_\mathrm{spin1},\phi_\mathrm{spin1},a_\mathrm{spin2},\cos\theta_\mathrm{spin2},\phi_\mathrm{spin2}\},
  \label{e:lambda}
\end{eqnarray}
where $\Mc = \frac{(M_1 M_2)^{3/5}}{(M_1 + M_2)^{1/5}}$ and $\eta = \frac{M_1 M_2}{(M_1 + M_2)^2}$ are  
the chirp mass and symmetric mass ratio, respectively; $d_\mathrm{L}$ is the luminosity distance to the source; $\phi_\mathrm{c}$ is an integration constant that specifies  
the GW phase at the time of coalescence $t_\mathrm{c}$, defined with respect to the centre of the Earth; $\alpha$ (right ascension)  
and $\delta$ (declination) identify the source position in the sky;  
%%the angles $\theta_{J_0}$ and $\phi_{J_0}$ (defined in the range $\theta_{J_0} \in \left[-\frac{\pi}{2},\frac{\pi}{2} \right]$  
%%and $\phi_{J_0}\in \left[0, 2\pi \right[$) identify the unit vector $\hat{\mathbf{J}}_0$; $d_\mathrm{L}$ is the  
 $\iota$ defines the inclination of the binary with respect to the line of sight; and $\psi$ is the polarisation angle of the waveform.
The spins are specified by $0 \le a_\mathrm{spin_{1,2}} \equiv S_{1,2}/M_{1,2}^2 \le 1$ as the dimensionless spin magnitude, and the angles 
$\theta_\mathrm{spin1,2}$,$\phi_\mathrm{spin1,2}$ for their orientations.

Given a network comprising $n_\mathrm{det}$ detectors, the data collected at the $a-$th  
instrument ($a = 1,\dots, n_\mathrm{det}$) is given by $x_a(t) = n_a(t) + h_a(t;\vec{\lambda})$,  
where $h_a(t;\vec{\lambda}) = F_{a,+}(t,\alpha,\delta,\psi)\,h_{a,+}(t;\vec{\lambda}) + F_{a,\times}(t,\alpha,\delta,\psi)\,h_{a,\times}(t;\vec{\lambda})$  
is the GW strain at the detector \citeaffixed{1994PhRvD..49.6274A}{see Eqs.\,2--5 in}
and $n_a(t)$ is the detector noise. The astrophysical signal  
is given by the linear combination of the two independent polarisations $h_{a,+}(t;\vec{\lambda})$  
and $h_{a,\times}(t;\vec{\lambda})$ weighted by the %\emph{time-dependent} 
antenna beam patterns $F_{a,+}(t,\alpha,\delta,\psi)$ and $F_{a,\times}(t,\alpha,\delta,\psi)$.

%\subsection{Waveform template in the ${3.5}$-pN order}

%Although the ${1.5}$-pN, simple-precession waveform is useful to investigate 
%the principal effects of spin on parameter estimation, a more realistic waveform is needed to analyse detected signals.
The waveform we use includes terms up to ${3.5}$-post-Newtonian (pN) order in phase and uses Newtonian amplitudes, with spin effects up to ${2.5}$-pN in phase.
We generate the waveform templates using the routine \texttt{LALGenerateInspiral()} with the approximant \texttt{SpinTaylor} from the injection package in the LSC Algorithm Library (LAL) \cite{LAL}, 
which closely follows the first section of \citename{2003PhRvD..67j4025B} \citeyear{2003PhRvD..67j4025B}.
%For comparison purposes we converted the usual set of parameters used in the LAL software 
%to the parameters in Eq.\,\ref{e:lambda}. In doing so, we fix 3 of the 15 parameters of the LAL parameter set, 
%setting the spin of the second member of the binary to be 0.
%An example of $h_a(t)$ for $a_\mathrm{spin}=0.5$ and $\theta_{SL}=20^\circ$ for both waveforms 
%is shown in figure%~\,\ref{fig:wave}. 
%Using the approximant \texttt{SpinTaylor}, this function calls \texttt{LALSTPNWaveform()} which does the 
%computation of the waveform. The projection on the interferometer arms is done using 
%\texttt{LALSimulateCoherentGW()} from the injection package, and the time delay is computed via 
%\texttt{LALTimeDelayFromEarthCenter()} from the same package.

%\begin{figure}  %For aastex
%%\begin{figure*} %For emulateapj
%  %\figurenum{}
%  %\epsscale{1.0}
%  %\plotone{f1.eps}
%  %\centering
%  \includegraphics[angle = 270, width=\linewidth]{detectorsignal.eps}
%  \caption{
%    {\bf (a)} Part of the ${1.5}$-pN time-domain waveform from a source with $a_\mathrm{spin}=0.5$ and  
%    $\theta_{SL}=20^\circ$.   
%    {\bf (b)} The ${3.5}$-pN waveform from a source with the same parameters. 
%    The waveforms start at 40\,Hz and are aligned at the coalescence time.
%    \label{fig:wave}
%  }
%\end{figure}  %For aastex
%%\end{figure*} %For emulateapj

\section{Parameter estimation: Methods}
\label{sec:methods}

In our Bayesian analysis we use MCMC methods to determine the multi-dimensional \emph{posterior} probability-density function (PDF) of the unknown parameter  
vector $\vec{\lambda}$ in equation~\ref{e:lambda}, given the data sets $x_a$ collected by a  
network of $n_\mathrm{det}$ detectors, a model $M$ of the waveform and the \emph{prior} $p(\vec{\lambda})$ on the  
parameters. Our priors are uniform in the parameters of Eq.\,\ref{e:lambda} (see \citeasnoun{2008CQGra..25r4011V} for details).
%\begin{eqnarray}
%\vec{\lambda} =& \{\Mc,\eta,\log{d_\mathrm{L}},t_\mathrm{c},\phi_\mathrm{c},\alpha,\cos\delta,\sin{\iota},\psi, \nonumber \\
%&a_\mathrm{spin1},\cos\theta_\mathrm{spin1},\phi_\mathrm{spin1},a_\mathrm{spin2},\cos\theta_\mathrm{spin2},\phi_\mathrm{spin2}\},
%\label{e:prior}
%\end{eqnarray}
One can compute the probability density via Bayes' theorem 
\be
p(\vec{\lambda}|x_a,M) = \frac{p(\vec{\lambda}|M) \, p(x_a|\vec{\lambda},M)}{p(x_a|M)}\,,
\label{e:jointPDF}
\ee
where 
%\be
%p(x_a|\vec{\lambda}) \propto 
%\exp\left\{-2
%\int_{f_l}^{f_h} \frac{\left| \tilde{x}_a(f) - \tilde{h}_a(f;\vec{\lambda})\right|^2}{S_a(f)}\,\mathrm{d}f
%\right\}
%\label{e:La}
%\ee
\be
\mathcal{L} \equiv p(x_a|\vec{\lambda},M) \propto 
\exp\left(
<x_a|h_a(\vec{\lambda})>-\frac{1}{2}<h_a(\vec{\lambda})|h_a(\vec{\lambda})>
\right)
\label{e:La}
\ee
is the \emph{likelihood function}, which measures how well the data fits the model $M$ for the parameter vector $\vec{\lambda}$. 
The term $p(x_a|M)$ is the \emph{marginal likelihood} or \emph{evidence}. In the previous equation
\be
<x|y>=4Re\left( \int_{f_{\rm low}}^{f_{\rm high}}\frac{\tilde{x}(f)\tilde{y}^{*}(f)}{S_a(f)}\,\mathrm{d}f \right)
\label{e:prod}
\ee
is the \emph{overlap} of signals $x$ and $y$, $\tilde x(f)$ is the Fourier transform of $x(t)$, and $S_a(f)$ is the noise power-spectral density in detector $a$.    
%For future reference, we also define match between two waveforms corresponding to different parameter values as the overlap between the normalised waveforms:
%\be
%M(h(\vec{\lambda_1}),h(\vec{\lambda_2}))=\frac{<h(\vec{\lambda_1})|h(\vec{\lambda_2})>}{\sqrt{<h(\vec{\lambda_1})|h(\vec{\lambda_1})><h(\vec{\lambda_2})|h(\vec{\lambda_2})>}} 
%\label{e:M}
%\ee
The likelihood computed for the injection parameters $\mathcal{L}_\mathrm{inj}=p(x_a|\vec{\lambda}_\mathrm{inj},M)$ 
is then a random variable that depends on the particular noise realisation $n_a$ in the data $x_a=h(\vec{\lambda}_\mathrm{inj})+n_a$. 
The injection parameters are the parameters of the waveform template added to the noise.
We define the signal-to-noise ratio (SNR) of the injection to be:
\be
\mathrm{SNR} = \frac{<x|h(\vec{\lambda}_{\rm inj})>}{\sqrt{<h(\vec{\lambda}_{\rm inj})|h(\vec{\lambda}_{\rm inj})>}}.
\ee
From here on, we use the expected value of the SNR, which is equal to the square root of twice the expectation value of $\log {\cal L_\mathrm{inj}}$:
\be
\mathrm{SNR} = \sqrt{<h(\vec{\lambda}_\mathrm{inj})|h(\vec{\lambda}_\mathrm{inj})>}.
\label{e:SNR}
\ee

To combine observations from a network of detectors with uncorrelated noise realisations (this is the case in this paper as we use two non-co-located detectors)
%we have $p(\vec{\lambda}|\{x_a; a = 1,\dots,n_\mathrm{det}\}) = \prod_{a=1}^{n_\mathrm{det}}\, p(\vec{\lambda}|x_a)\,.$
we have the likelihood $p(\vec{x}|\vec{\lambda},M) = \prod_{a=1}^{n_\mathrm{det}}\, p(x_a|\vec{\lambda},M)\,$, for $\vec{x} \equiv \{x_a: a = 1,\dots,n_\mathrm{det}\}$ and
\be
p(\vec{\lambda}|\vec{x},M) = \frac{p(\vec{\lambda}|M)\, p(\vec{x}|\vec{\lambda},M)}{p(\vec{x}|M)}.
\label{e:Bayes}
\ee

The numerical computation of the %joint and \emph{marginalised} 
PDF involves the evaluation of a large multi-modal, multi-dimensional integral. Markov-chain Monte-Carlo  
(MCMC) methods \citeaffixed[and references therein]{gilks_etal_1996,gelman_etal_1997}{\emph{e.g.}}
have proved to be especially effective in tackling this numerical problem.
We developed an adaptive \citeaffixed{figueiredo_jain_2002,atchade_rosenthal_2005}{see}
MCMC algorithm to explore the parameter space $\Lambda$ efficiently while requiring the least amount of tuning for the specific signal analysed; the code is an extension of the one developed by some of the authors to explore  
MCMC methods for binaries without spin \cite{2006CQGra..23.4895R,2007PhRvD..75f2004R}.
We implemented parallel tempering \cite{1996JPSJ...65.1604H,1997CPL...281..140H,2007PhD..Auck} to improve the sampling.   
It consists of running several MCMC chains in parallel, each with a different ``temperature'', which 
can swap parameters under certain conditions. Only the $T=1$ chain is currently used for post-processing.

In Eq.\,\ref{e:Bayes} we applied Bayes' theorem to obtain the probability of a specific parameter vector value ($\vec{\lambda}$) given the observed data $\vec{x}$ and the model $M$. 
The theorem can also be applied to compute the probability of a specific \emph{model} $M_i$ given the observed data:
\be
p(M_i|\vec{x}) = \frac{p(M_i)\, p(\vec{x}|M_i)}{p(\vec{x})}.
\label{e:BayesModel}
\ee
We compare the two models $M_i$ and $M_j$ by computing the \emph{odds ratio}:
\be
O_{i,j}= \frac{p(M_i|\vec{x})}{p(M_j|\vec{x})} = \frac{p(M_i)\, p(\vec{x}|M_i)}{p(M_j)\, p(\vec{x}|M_j)} = \frac{p(M_i)}{p(M_j)} B_{i,j},
\label{e:Odds}
\ee
where 
\be
B_{i,j} = \frac{p(\vec{x}|M_i)}{p(\vec{x}|M_j)} 
\label{e:BayesRatio}
\ee
is the \emph{Bayes factor} of the two models, and we recognise the evidence $p(\vec{x}|M_i)$ from Eq.\,\ref{e:Bayes}. 
The evidence must be marginalised over the parameters of the model in order to compute the Bayes factor:
\be
%Z_i = p(\vec{x}|M_i) = \int_{\Lambda} p(\vec{\lambda}|M_i) \, p(\vec{x}|\vec{\lambda},M_i) \,\mathrm{d}\vec{\lambda}.
p(\vec{x}|M_i) = \int_{\Lambda} p(\vec{\lambda}|M_i) \, p(\vec{x}|\vec{\lambda},M_i) \,\mathrm{d}\vec{\lambda}.
\label{e:evidence}
\ee
There are existing algorithms dedicated to the computation of this integral, and of the Bayes factor.  For instance, \emph{nested sampling} \cite{MR2282208} has been shown 
to be very efficient in the case of non-spinning gravitational-wave sources \cite{2009arXiv0911.3820V}, and can in addition be used to produce PDFs of the parameters. 
As a by-product of the exploration of the parameter space with MCMC, it is possible to compute the evidences of the models used. We have implemented the harmonic-mean method \cite{1994JSTOR..N},
in which the evidence is approximated by:
\be
%Z_i\,\approx\,\sum_{k=1}^N p(\vec{\lambda}_k|M_i)\,p(\vec{x}|\vec{\lambda}_k,M_i) \,V_{\vec{\lambda}_k},
p(\vec{x}|M_i)\,\approx\,\sum_{k=1}^N p(\vec{\lambda}_k|M_i)\,p(\vec{x}|\vec{\lambda}_k,M_i) \,V_{\vec{\lambda}_k},
\ee
where $\{\vec{\lambda}_k: k = 1,\dots,N\}$ is the set of $N$ points sampled by the MCMC, and $V_{\vec{\lambda}_k}$ 
is the volume of parameter space associated with the point $\vec{\lambda}_k$. Since the MCMC algorithm samples according to the posterior 
(and, up to a proportionality constant, converges towards posterior PDF), the density of points in the chain at a certain location $\vec{\lambda}_k$ 
in the parameter space $\Lambda$ will become proportional to the posterior for large $N$.  It follows that
\be
\lim_{N \to \infty}\,V_{\vec{\lambda}_k} = \frac{\alpha_i}{p(\vec{\lambda}_k|M_i)\,p(\vec{x}|\vec{\lambda}_k,M_i)},
\ee
with $\alpha_i$ a proportionality constant. We then have %$Z_i\,\approx\,\sum_{k=1}^N \alpha_i = N\,\alpha_i$, and obtain the estimate for $\alpha_i$ by considering the whole prior volume $V_t$:
$p(\vec{x}|M_i)\,\approx\,\sum_{k=1}^N \alpha_i = N\,\alpha_i$, and obtain the estimate for $\alpha_i$ by considering the whole parameter space volume $V_t$:
\be
V_t\,\approx\,\sum_{k=1}^N V_{\vec{\lambda}_k} = \sum_{k=1}^N \frac{\alpha_i}{p(\vec{\lambda}_k|M_i)\,p(\vec{x}|\vec{\lambda}_k,M_i)}.
\ee
Finally,
\be
%Z_i\,\approx\,N\,V_t\,\left[ \sum_{k=1}^N \frac{1}{p(\vec{\lambda}_k|M_i)\,p(\vec{x}|\vec{\lambda}_k,M_i)} \right]^{-1},
p(\vec{x}|M_i)\,\approx\,N\,V_t\,\left[ \sum_{k=1}^N \frac{1}{p(\vec{\lambda}_k|M_i)\,p(\vec{x}|\vec{\lambda}_k,M_i)} \right]^{-1},
\ee
which is the harmonic mean of the posterior values sampled by the MCMC. The issue with this method is that it gives too much weight to low-posterior points, 
which lie in a part of the parameter space that is badly sampled, by design, by the MCMC. 
The estimate of the evidence is then very sensitive to the quality of the sampling of a particular run. We are looking into other algorithms in order to remedy this problem, 
\emph{e.g.}\ by using the higher-temperature chains produced by parallel tempering \cite{earl-2005} (we currently use the $T=1$ chain only), or by using a well sampled subset of points \cite{vanhaasteren-2009} to estimate the probability constant $\alpha_i$. A summary of the methods used in our MCMC code was published in \citeasnoun{2008CQGra..25r4011V};
a more complete technical description of the \textsc{SPINspiral} code will be available in \cite{vandersluysprep}.

\section{Parameter estimation: Results}
\label{sec:results}

%In this section, we present the results of our analysis of a number of different example data sets with \textsc{SPINspiral}.

\subsection{Data sets}
\label{sec:data}

For these proceedings, we analyse three different data sets, each containing the data for the 4-km LIGO detectors at Hanford (H1)
and Livingston (L1):
\begin{description}
\item[DS1:] a coherent software injection with a total SNR of 11.3 into synthetic Gaussian, stationary noise, simulated for the H1 and L1 detectors;
\item[DS2:] a coherent software injection of the same signal, with a total SNR of 11.3, into ``quiet'' LIGO detector data from H1 and L1;
\item[DS3:] raw LIGO data from H1 and L1, containing a known, coincident glitch of seismic origin, with a total SNR of 11.3.
\end{description}

For the data sets DS1 and DS2, the injected signal is that of a $10\,M_\odot$ spinning BH and a $1.4\,M_\odot$ spinning NS in an inspiralling 
binary system. A low-mass Compact Binary Coalescence Group search \cite{2009PhRvD..79l2001A} does not produce a GW trigger for the data segment DS2; hence we designate it ``quiet''.
%The ``quiet'' data for DS2 indicates that there is no GW trigger from the low mass Compact Binary Coalescence Group search \cite{2009PhRvD..79l2001A} in these data. 
The distance of each of the injections is scaled to obtain an SNR of 11.3, equal to that of the 
glitch in DS3, but computed with different waveforms: a SpinTaylor waveform (see section\,\ref{sec:GW}) for DS1 and DS2, and a non-spinning, ${2}$-pN waveform (see section\,\ref{sec:glitch}) for DS3.
The other parameters of the injection are:
\begin{eqnarray}
  \vec{\lambda} =& \{\Mc=2.99\,M_\odot,\eta=0.107,d_\mathrm{L},t_\mathrm{c},\phi_\mathrm{c}=85.9^\circ,\alpha=17.4\,h,\delta=61.6^\circ, \nonumber \\
  &i=52.8^\circ,\psi=11.6^\circ,a_\mathrm{spin1}=0.6,\theta_\mathrm{spin1}=78.5^\circ,\phi_\mathrm{spin1}=63.0^\circ, \nonumber \\
  &a_\mathrm{spin2}=0.4,\theta_\mathrm{spin2}=120.0^\circ,\phi_\mathrm{spin2}=315.1^\circ\},
  \label{e:parameters}
\end{eqnarray}
where we assigned a spin of 0.4 to the neutron star, which is higher than astrophysically plausible, for testing purposes only. 
In DS3, no signal is injected.  For our analyses, we use the data of both 4-km LIGO detectors H1 and L1.

%Here we present results obtained by injecting a signal into simulated and real interferometer noise,
%and computing the posterior PDFs with MCMC techniques, for a fiducial  
%source consisting of a $10\,M_\odot$ spinning BH and a $1.4\,M_\odot$ spinning NS in a binary system at a distance scaled to give an SNR of 11.3.
%We also ran the MCMC algorithm on simulated noise, real noise and real and glitchy noise without injecting a signal.
%\ref{sec:comparison} for parameters values). We consider a number of cases for which we change the BH spin parameters. 

\subsection{MCMC simulations}
\label{sec:runs}

The MCMC analysis that we carry out on each data set consists of 10 independent Markov chains,
each with a length of about a million iterations and composed of 5 chains at different temperatures 
for parallel tempering. From now on, we will refer to the $T=1$ chain as \emph{the chain}, since the hotter chains were not used in the post-processing.
%%depending on the number of detectors and the waveform version.
%%We kept the CPU time the same for each case, resulting in fewer iterations for a higher number of detectors and/or a higher pN order.
The part of the chains that is analysed is that after the \emph{burn-in} period \citeaffixed{gilks_etal_1996}{see \emph{e.g.}},
the length of which is determined
automatically as follows: we determine the absolute maximum likelihood $\log({\cal L}_\mathrm{max})$, 
defined as the highest value for $\log[p(\vec{x}|\vec{\lambda},M)]$ 
obtained over the ensemble of parameter sets $\vec{\lambda}$ in any of our individual Markov chains. Then for
each chain we include all the iterations \emph{after} the chain reaches a likelihood value of 
$\log({\cal L}_\mathrm{max})-2$ for the first time.
This results in a convergence test as well, since some of the independent chains may not reach this threshold value. Typically,
we demand that more than 50\% of our chains meet this condition before we consider the MCMC run as \emph{converged}, although
we consider results as \emph{robust} if they have a convergence rate of 80\% or more. This convergence test is a measure of the quality of our sampling in a given number of iterations. 
All our Markov chains start at values that are randomly offset from the injection values. The starting values for 
$\Mc$ and $t_\mathrm{c}$ are drawn from a Gaussian distribution 
centred on the injection value, with a standard deviation of $0.025\,M_\odot$ and 
10\,ms respectively. In real analysis, the two Gaussian distributions are centred on the values from the template bank based search of the Compact Binary Coalescence group \cite{2009PhRvD..79l2001A} which will have triggered the MCMC followup.
%, which is much wider than the accuracy of a typical detection trigger, allowing us to conduct the convergence test described above.
The other thirteen parameters are drawn uniformly from their allowed ranges.
\textsc{SPINspiral} needs to run for typically a few days in order to show the first results and a week or two
to accumulate a sufficient number of iterations for good statistics, each chain using
a single 2.8\,GHz CPU.

%\subsection{Synthetic Gaussian noise and real detector data}
\subsection{Analysis of data sets DS1 and DS2}
\label{sec:subresults}

\begin{figure}%
\centering 
\includegraphics[width=.7\textwidth,angle=270]{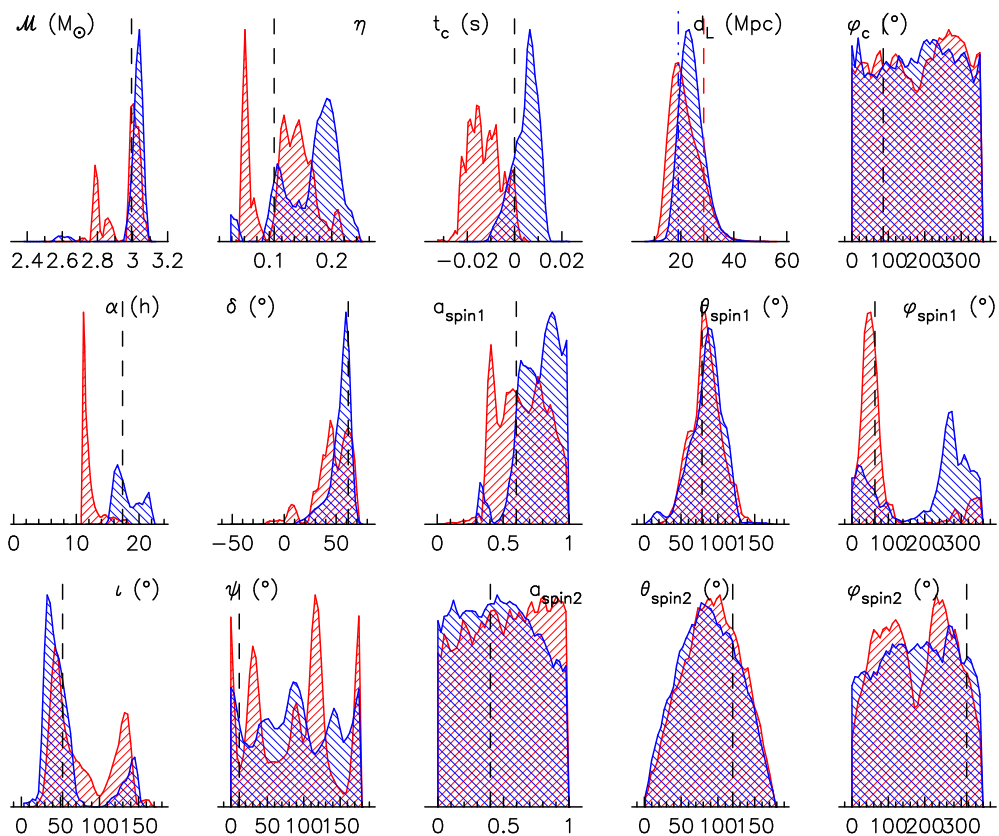}
%\subfloat[One dimensional PDFs]{\label{subfig:gaussian1d}\includegraphics[width=.5\textwidth,angle=270]{comp_pdfs/comp_pdfs__example__1d.eps}} \\
%\subfloat[Mc-eta, synthetic]{\label{subfig:gaussianMcetaInj}\includegraphics[width=.30\textwidth]{gaussianinj/GPS0894377000_H1L1__SpinTaylor15_3.5pN_2sp__pdf2d__Mc-eta.eps}}
%\subfloat[Mc-eta, detector]{\label{subfig:gaussianMcetaNoise} \includegraphics[width=.30\textwidth]{detectorinj/GPS0846226300_H1L1__SpinTaylor15_3.5pN_2sp__pdf2d__Mc-eta.eps}}\\

%
%\subfloat[Skymap, synthetic]{\label{subfig:gaussianSkymapInj} \includegraphics[width=.45\textwidth]{gaussianinj/GPS0894377000_H1L1__SpinTaylor15_3.5pN_2sp__pdf2d__RA-dec.eps}}
%\subfloat[Skymap, detector]{\label{subfig:gaussianSkymapNoise} \includegraphics[width=.45\textwidth]{detectorinj/GPS0846226300_H1L1__SpinTaylor15_3.5pN_2sp__pdf2d__RA-dec.eps}}

\caption{
  One-dimensional marginalised PDFs for all 15 parameters from our analysis of data sets DS1 (hatched upward; red in the online colour version)
  and DS2 (hatched downward; blue in the online colour version). The vertical dashed lines mark the injection values.
}
\label{fig:PDFs} 
\end{figure}

We analysed the data sets DS1 and DS2 as described in section~\ref{sec:data} and the results of both analyses passed the convergence
test described in section~\ref{sec:runs} with convergence rates of 70\% and 80\%, respectively.
The resulting one-dimensional marginalised PDFs from both analyses are shown in figure~\,\ref{fig:PDFs}.  
%The results from DS1 are hatched upward (red in the colour version), those of DS2 are hatched downward (blue).

Table~\,\ref{table:numbers} shows the median and the width of the 95\%-probability ranges for each parameter. 
The differences we find between the results for DS1 and DS2 may be attributed to the particular noise realisations in this example,
and most parameters yield similar PDFs and accuracies.

\begin{table}
  \caption{
    Median and width of the 95\%-probability ranges for each parameter of the analyses of data sets DS1 and DS2.
    The column \emph{recovered} indicates whether or not the 95\% range includes the injection value.
    %\vivien{the difference in accuracy in chirp mass is expected to disappear when I update the numbers}
    \label{table:numbers}
  }
  \begin{indented}
  \item[] \begin{tabular}{l | l | lll | lll}
      \br
    							&				& \multicolumn{3}{c|}{DS1 (synthetic noise)}                       & \multicolumn{3}{c}{DS2 (detector noise)}   \\
    							&	injection		&	median                 &    95\% width    & recovered & median          &    95\% width & recovered \\
    \mr
$\Mc\,(M_\odot)$                   		&	2.99		&	3.006	&	0.294	& yes	&	3.041	&	0.122	& yes	\\
$\eta$                                      		&	0.107	&	0.133	&	0.145	& yes	&	0.183	&	0.144	& yes	\\
$d_\mathrm{L}$\,(Mpc)        		&	28.615	&	21.240	&	20.764	& yes	&	24.144	&	17.238	& yes	\\
$t_\mathrm{c}$\,(s)                 		&	0.000	&	-0.013	&	0.024	& yes	&	0.006	&	0.019	& yes	\\
$\phi_\mathrm{c}\,(^\circ)$     		&	85.944	&	189.745	&	342.398	& yes	&	185.482	&	343.175	& yes	\\
$\alpha$\,(h)                           		&	17.380	&	11.684	&	5.349	& \textbf{no}		&	17.786	&	6.320	& yes	\\
$\delta\,(^\circ)$                          		&	61.642	&	49.326	&	64.346	& yes	&	58.390	&	39.796	& yes	\\
%$Sky area deg^2$				&	---		&	---	          &      799            & yes	&       ---              &       1151         & yes         \\
$i\,(^\circ)$       					&	52.753	&	67.056	&	110.735	& yes	&	46.850	&	122.787	& yes	\\
$\psi\,(^\circ)$        				&	11.459	&	93.162	&	176.358	& yes	&	88.706	&	173.869	& yes	\\
$a_\mathrm{spin1}$      			&	0.600	&	0.658	&	0.594	& yes	&	0.804	&	0.478	& yes	\\
$\theta_\mathrm{spin1}\,(^\circ)$  	&	78.463	&	85.490	&	83.110	& yes	&	89.225	&	85.787	& yes	\\
$\phi_\mathrm{spin1}\,(^\circ)$       	&	63.025	&	57.171	&	335.592	& yes	&	263.014	&	345.700	& yes	\\
$a_\mathrm{spin2}$      			&	0.400	&	0.532	&	0.945	& yes	&	0.475	&	0.940	& yes	\\
$\theta_\mathrm{spin2}\,(^\circ)$  	&	120.000	&	94.687	&	150.544	& yes	&	89.406	&	146.101	& yes	\\
$\phi_\mathrm{spin2}\,(^\circ)$       	&	315.127	&	181.959	&	327.603	& yes	&	184.681	&	339.071	& yes	\\
$M_1\,(M_\odot)$        			&	10.002	&	8.533	&	8.849	& yes	&	6.421	&	6.536	& yes	\\
$M_2\,(M_\odot)$        			&	1.400	&	1.598	&	1.277	& yes	&	2.036	&	1.564	& yes	\\

    \br
  \end{tabular}
  \end{indented}
\end{table}

The PDFs of the parameters that describe the spin of the NS follow the prior distributions in both runs. This justifies ignoring the NS spin (by fixing $a_\mathrm{spin2}$ to 0.0 in the recovery template)
for this mass ratio \cite{2008ApJ...688L..61V}. %However, the additional degrees of freedom provided by the second spin yield more accurate 
%measurements in some of the other parameters, \emph{e.g.}\ for $\Mc$ our 95\% range is 0.294 with second spin included, and 0.322 when ignoring the second spin (using data set DS1).
For each of the two data sets, DS1 and DS2, we computed the Bayes factor to compare the evidence for the following two models: 
$M_1$: a ${3.5}$-pN inspiral waveform embedded in Gaussian noise, and $M_2$: Gaussian noise only.
The values are listed in table~\,\ref{table:Bayes}. In both cases, the Bayes factor is large, providing strong evidence for a GW
signal in the data. The difference in Bayes factor between DS1 and DS2 is attributed to an inherent spread due to different noise realisations, 
and the uncertainties of our method to estimate the Bayes factor (section~\ref{sec:methods}). 
The results in this section show an illustrative example, but cannot be used to draw firm conclusions.  However, it is clear that they warrant
a larger, systematic study of these phenomena with the methods described here.

\begin{table}
  \caption{
    Bayes factors $B_{1,2}$ between the models $M_1$: a ${3.5}$-pN inspiral waveform embedded in Gaussian noise, and $M_2$: Gaussian noise only (section~\ref{sec:subresults}) for data sets DS1 and DS2 (see section~\ref{sec:data}).
    \label{table:Bayes}
  }
  \begin{indented}
  \item[] \begin{tabular}{ccccc}
      \br
      &  DS1 (Gaussian noise)  & DS2 (detector data)  & DS3 (glitch)\\
      \mr
      $\log_e B_{1,2}$   & 52.9 & 43.5 & 68.5  \\    
      \br
    \end{tabular}
  \end{indented}
\end{table}

\subsection{Analysis of data sets DS2 and DS3}
\label{sec:glitch}

\begin{figure}
  \centering 
  \includegraphics[width=.85\textwidth,angle=0]{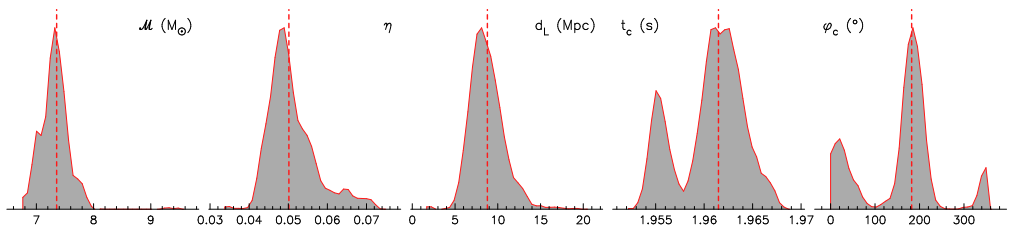}
%  \subfloat[DS3 (glitch), one-dimensional  PDFs]{\label{subfig:glitch1d}\includegraphics[width=.8\textwidth,angle=0]{glitch/GPS0846471912_H1L1__SpinTaylor15_3.5pN_2sp__pdfs.eps}} \\
%  \subfloat[DS2 (injection), sky map]{\label{subfig:goodSkymap}\includegraphics[width=.4\textwidth]{detectorinj/GPS0846226300_H1L1__SpinTaylor15_3.5pN_2sp__pdf2d__RA-dec.eps}}
%  \qquad
%  \subfloat[DS3 (glitch), sky map]{\label{subfig:glitchSkymap}\includegraphics[width=.4\textwidth]{glitch/GPS0846471912_H1L1__SpinTaylor15_3.5pN_2sp__pdf2d__RA-dec.eps}}
  \caption{
    One-dimensional marginalised PDFs of a few selected parameters from our analysis of data set DS3.
    The vertical dashed lines indicate the median of each PDF.
    %Sky map from the analysis of data set DS2 (\ref{subfig:goodSkymap}) and from that of DS3, (\ref{subfig:glitchSkymap}).
    %The grey scales or colours show the different probability areas (1-$\sigma$, 2-$\sigma$ and 3-$\sigma$ as indicated at the top of each panel).
  }
  \label{fig:glitch} 
\end{figure}

On November 2nd 2006, seismic activity at Hanford and Livingston resulted in a coincident ``glitch'' in the data from 
the H1 and L1 LIGO detectors. 
These glitches were recovered by the Compact Binary Coalescence detection pipeline at an SNR of 11.3, using non-spinning, 
stationary-phase-approximation templates, Newtonian in amplitude and 2.0-pN in phase \cite{2009PhRvD..79l2001A}.
We defined the corresponding data set as DS3 in section~\ref{sec:data} and analysed the data as if it had yielded a GW trigger.
The convergence test from section~\ref{sec:runs} yields a 20\% convergence rate, which results in our rejection of the 
results as \emph{not converged}. 
However, when we nevertheless construct the marginalised one-dimensional  PDFs from the data of the two converged chains
(because of the small number of data points, the resulting PDFs may not be very accurate),
they are similar in appearance to those from DS2 (see figure~\,\ref{fig:glitch}). 
The Bayes factors in table~\,\ref{table:Bayes} even suggest that the data set DS3 is more consistent with containing a GW signal than DS2 (with the caveat that the SNRs of DS2 and DS3 were not computed the same way).
%The marginalised one-dimensional  PDFs have the same features as an injection (in terms of width, number of modes...). 
%And the Bayes factor in table~\,\ref{table:Bayes} is consistent with an gravitational wave signal. 
%The sky map for DS2 in figure~\,\ref{subfig:goodSkymap} displays the expected sky ring for an analysis using two non-colocated detectors, 
On the other hand, the low value for the median of $\eta$ (0.05) corresponds to a mass ratio of 18, which is near the limit of the regime where post-Newtonian expansions are valid.
In particular, a small value for eta suggests a slow frequency evolution which may indicate a spike in the frequency spectrum that dominates the signal.
In addition, we find that the sky map for DS3 does not display the (parts of a) sky ring that is expected for an analysis using two non-co-located detectors \citeaffixed{2009CQGra..26k4007R}{see \textit{e.g.}}.
%This is just simple example, and the issue requires deeper investigations with a number of different glitches. 
These results indicate that we should thoroughly verify our tests, such as the convergence criterion described here, using a 
large number of different glitches.

\section{Conclusions}
\label{sec:concl}

We have developed the code \textsc{SPINspiral} which can do a complete parameter analysis of the gravitational-wave signals from 
quasi-circular compact-binary inspirals.  We presented an example of the analysis of software injections into both simulated Gaussian 
noise (DS1) and LIGO-detector data (DS2). We also presented an analysis of a data set containing no injection, but a ``glitch'' coincident 
in two LIGO interferometers (DS3). These examples demonstrate a remarkable similarity between the results obtained
from a GW signal injected in Gaussian noise and a similar signal in detector data. The Bayes factors are also similar, 
%where we note that our Bayes factors are computed from data that was sampled sub-optimally for this purpose.
where we note that our present technique for computing the Bayes factor yields estimates with significant variance, and more precise estimates should be possible in the future.
%with the caveat that our Bayes factor computation has a non-negligible variance and needs to be improved. 
In addition, we find that although the Markov chains in the analysis of a coincident glitch in LIGO data do not converge, 
the resulting PDFs could look remarkably consistent with a simulated GW signal. 
We plan to run our code on a very large number of coincident triggers from the LIGO Compact Binary Coalescence search pipeline 
(noise events that are somehow being registered as resembling a binary inspiral) in order to get a good sense of how to distinguish them from actual inspirals.
% some of its behaviour on real data. Further investigations are warranted to determine whether real detector noise really allows similar accuracies than synthetic Gaussian noise, and whether 
%glitches (which by nature are of all different kinds) pose a problem.
We conclude that further, detailed investigations are necessary to ensure we can rely on the robustness of our tests.

\ack
The authors acknowledge 
a CITA National Fellowship to the UoA for MvdS, 
the NSF astronomy and astrophysics postdoctoral fellowship under the award AST-0901985 to IM, 
a NSF Gravitational Physics grant (PHY-0854790) to NC and 
the Max-Planck-Society (CR). 
Computations were performed on the Fugu computer cluster funded by NSF MRI grant PHY-0619274 to VK. 
%\todo{Please send me your acknowledgements...}

\section*{References}
\bibliographystyle{jphysicsB}
\bibliography{NRDA09}{}

\end{document}